\documentclass[12pt]{article}
\begin{document}
\begin{center}
{\bf FUNCTIONAL APPROACH TO PHASE SPACE FORMULATION
                     OF QUANTUM MECHANICS}\\
                       {A.K.Aringazin\\
               Department of Theoretical Physics,
                  Karaganda State University,
                 Karaganda 470074, Kazakstan\\
{\tt ascar\mbox{@}ibr.kargu.krg.kz}\\[1cm]
{\small\it Proc. of the XVIII Workshop on High Energy Physics and
Field Theory, 26-30 June 1995, Protvino, Russia, Eds.V.A.Petrov,
A.P.Samokhin, R.N.Rogalyov (IHEP, Protvino, 1996) pp. 322-327.}
}\\[1cm]
\end{center}

{We present BRST gauge fixing approach to quantum
mechanics in phase space.
 The theory is obtained by $\hbar$-deformation of the
cohomological classical mechanics described by $d=1$, $N=2$ model.
 We use the extended phase space supplied by the
path integral formulation with $\hbar$-deformed symplectic structure.}\\

\section{Introduction}\label{QS1}

 Recently, in a series of papers Gozzi, Reuter and Thacker\cite{GRT}
developed path integral approach to classical mechanics.
 The physical states of the theory has been
analyzed\cite{GRT,Aringazin-93-HJ,Aringazin-93-PL}.
 This theory can be viewed as one-dimensional
cohomological field theory, in the sense that the resulting
BRST exact Lagrangian is derived by fixing symplectic diffeomorphism
invariance of the zero underlying Lagrangian.
 The BRST formulation of the cohomological classical mechanics
has been given\cite{Aringazin-94-HJ}, and machinery of modern
topological quantum field theories has been
used\cite{Aringazin-95-HJ1,Aringazin-95-HJ2}
to analyze the physical states, associated $d=1$, $N=2$ supersymmetric
model, BRST invariant observables, and correlation functions.
 This field theoretic description provides a powerful tool
to investigate the properties of Hamiltonian systems,
such as ergodicity, Gibbs distribution, and Lyapunov exponents.

 Further development can be made along the line of the phase
space formulation of ordinary quantum mechanics originated by
Weyl, Wigner and Moyal\cite{Weyl-Wigner-Moyal}.
 The key point one could exploit here is that it is
treated\cite{JP-75}-\cite{Aringazin-94-JP} as a {\sl smooth
$\hbar$-deformation} of the classical mechanics.
 Indeed, there is an attractive possibility to give an explicit
geometrical BRST formulation of the model describing quantum
mechanics in phase space.
 The resulting theory could be thought of as a topological phase
of quantum mechanics in phase space.
 The crucial part of the work has been done by Gozzi and
Reuter\cite{GR-92d} in the path integral formulation, where the
associated extended phase space
and quantum $\hbar$-deformed exterior differential calculus in
quantum mechanics has been proposed.
 The core of this formulation is in the deforming of the Poisson
bracket algebra of classical observables.

 The central point we would like to use here is that the extended
phase space can be naturally treated as the cotangent superbundle
$M^{4n|4n}$ over $M^{2n}$ endowed with the second symplectic
structure $\Omega$ and (graded) Poisson brackets (\ref{epb}).
 Besides clarifying the meaning of the ISp(2) group algebra, appeared
as a symmetry of the field theoretic model, it allows one,
particularly, to combine symplectic geometry and techniques of fiber
bundles.
 The underlying reason of our interest in elaborating the fiber
bundle construction is that one can settle down Moyal's
$\hbar$-deformation in a {\sl consistent} way by using both of the
Poisson brackets, $\{,\}_\omega$ and $\{,\}_\Omega$. Namely, the
two symplectic structures and Hamiltonian vector fields coexisting
in the single fiber bundle are related to each other.
 Note that this relation is not direct since
$\{a^i,a^j\}_\omega=\omega^{ij}$ while for the projection of
coordinates in the fundamental Poisson bracket
$\{\lambda^a,\lambda^b\}_\Omega=\Omega^{ab}$ to the base $M^{2n}$
we have $\{a^i,a^j\}_\Omega=0$.
 Also, $Z_2$ symmetry of the undeformed Lagrangian
can be used as a further important requirement for the
deformed extension.
 Naively, the problem is to construct $\hbar$-deformed BRST exact
Lagrangian, identify BRST invariant observables, and study BRST
cohomology equation and corresponding correlation functions.
 Also, having the conclusion that the $d=1$, $N=2$ supersymmetry
plays so remarkable role in the classical case it would be
interesting to investigate its role in the quantum mechanical case.

 In this way, one might formulate, particularly, quantum analogues
of Lyapunov exponents in terms of correlation
functions rather than to invoke to nearby trajectories, which make
no sense in quantum mechanical case. The case of compact classical
phase space corresponds to a finite number of quantum states.
 Also, we note that for chaotic systems expansion on the periodic
orbits constitutes the only semiclassical quantization scheme known.
Perhaps, this is a most interesting problem, in view of the recent
studies of quantum chaos.

 However, we should to emphasize here that the geometrical BRST
analogy with the classical case is not straightforward, as it may
seem at first glance, since one deals with non-commutative
geometry\cite{Connes-93} of the phase space in quantum mechanical
case (see Ref.\cite{Aringazin-94-JP} and references therein).
 Particularly, quantum mechanical observables of interest are
supposed to be analogues of the closed $p$-forms on $M^{2n}$,
with non-commuting coefficients arising to non-abelian cohomology.

 In this paper, we present BRST gauge fixing approach to quantum
mechanics in phase space.

 The paper is organized as follows.
In Sec.~\ref{QS2}, we briefly present the BRST construction
of the $d=1$, $N=2$ model characterizing classical dynamical
systems.
In Sec.~\ref{QS3}, we introduce symplectic structure on the
extended phase space naturally supplied by the field content.
In Sec.~\ref{QS4}, we make the $\hbar$-deformation of the
symplectic structures using graded Moyal's brackets.

\section{The BRST formulation}\label{QS2}

 Starting point is the partition function\cite{Aringazin-94-HJ}
\begin{equation}
Z = \int Da\, \exp\, iI_{0},                      \label{Z0}
\end{equation}
where $I_{0} = \int dt {\cal L}_{0}$, and the Lagrangian
is trivial, ${\cal L}_{0} = 0$.
 The basic field $a^i(t)$ is the map from one-dimensional
space $M^1$ to $2n$-dimesional symplectic manifold $M^{2n}$.
 The BRST fixing of the symplectic diffeomorphism invariance
yields\cite{Aringazin-94-HJ}
\begin{equation}
Z = \int DX\, \exp\, iI,                           \label{Z}
\end{equation}
where $I = I_{0} + \int dt {\cal L}_{gf}$, and
$DX$ represents path integral over the fields entering the
gauge fixing Lagrangian ${\cal L}_{gf}$.
 The Hamilton function ${\cal H}$ associated to
${\cal L}_{gf}$ has been found in the form
\begin{equation}
{\cal H}  = q_{i}h^{i}
          + i\bar c_{i}\partial_{k}h^{i}c^{k}
          - \alpha a^{i}q_{i}
          - i\alpha c^{i}\bar c_{i},                   \label{H}
\end{equation}
where $c^{i}$ and $\bar c_{i}$ are ghost and antighost fields,
respectively, $q_{i}$ is a Lagrange multiplier, $h^i$ is
Hamiltonian vector field, and $\alpha$ is a real parameter.
 The Hamilton function (\ref{H}) is BRST and anti-BRST exact.
 In the following, we use the delta function gauge by putting
$\alpha = 0$ in (\ref{H}).
 We refer the reader to Refs.
\cite{Aringazin-94-HJ,Aringazin-95-HJ1,Aringazin-95-HJ2}
for precise and detailed
development of cohomological classical mechanics
and geometrical meaning of the fields $q_{i}$, $c^{i}$ and $\bar c_{i}$.

\section{Symplectic structures}\label{QS3}

In order to prepare for the $\hbar$-deformation,
we need to identify geometrically the phase space and
algebra of observables supplied by the BRST procedure.

 Symplectic manifold $M^{2n}$ can be viewed as cotangent fiber bundle
$(M^{2n}, M^{n},$ $T^{*}_{x}M^{n},\omega)$
with the base space $M^{n}$,
fiber $T^{*}_{q}M^{n}$, and fundamental symplectic two-form $\omega$,
which in local coordinates,
$a^{i}= (p_1,\dots ,p_n , x^1,\dots ,x^n)$,
$a \in M^{2n}$,
$x \in M^{n}$,
$p \in T^{*}_{x}M^{n}$, is
$\omega = \frac{1}{2}\omega_{ij}da^{i}\wedge da^{j}$;
$\omega_{ij} =-\omega_{ji}$, $\omega_{ij}\omega^{jm} = \delta^{i}_{m}.$
 In Hamiltonian mechanics, $M^{2n}$ plays the role of phase space
equipped by standard Poisson brackets,
$
\{ f,g\}_{\omega} = f\bar\partial_{i}\omega^{ij}\vec\partial_{j}g.
$
We assume $\omega_{ij}$ to be a constant matrix, {\sl i.e.}
use Darboux coordinates.

 Taking the phase space $M^{2n}$ as a base space, we consider
a cotangent fiber bundle $M^{4n}$ over it,
$(M^{4n}, M^{2n}, T^{*}_{a}M^{2n}, \Omega )$,
where two-form $\Omega$ defines symplectic structure on $M^{4n}$, and
is assumed to be closed, $d\Omega = 0$, and non-degenerate.
In local coordinates on $M^{4n}$,
$\lambda^{a}= (q_1,\dots ,q_{2n} , a^1,\dots ,a^{2n})$,
$\lambda \in M^{4n}$,
$a \in M^{2n}$,
$q \in T^{*}_{a}M^{2n}$,
the two-form $\Omega$ is represented as
$\Omega = \frac{1}{2}\Omega_{ab}d\lambda^{a}\wedge d\lambda^{b}$;
$\Omega_{ab} =-\Omega_{ba}$, $\Omega_{ab}\Omega^{bc} = \delta^{a}_{c}.$

 The cotangent bundle $M^{4n}$ can be thought of as a second generation
phase space equipped by the Poisson brackets,
\begin{equation}
                                                           \label{epb}
\{ F, G\}_{\Omega} = F\bar\partial_{a}\Omega^{ab}\vec\partial_{b}G,
\end{equation}
where $\partial_{a} =\partial /\partial\lambda^{a}$,
in view of the sequence
\begin{equation}
                                                     \label{maps}
M^{n} \stackrel{p^{-1}_{0}}{\rightarrow}
                (M^{2n}, M^{n}, T^{*}_{x}M^{n}, \{ ,\}_{\omega})
      \stackrel{p^{-1}}{\rightarrow}
                (M^{4n}, M^{2n}, T^{*}_{a}M^{2n}, \{ ,\}_{\Omega} )
\end{equation}
 Natural projection $p$ is provided by $p:\ (q,a) \mapsto (0,a)$.

 The ghost and antighost fields, as Grassmannian variables,
can be naturally added to the symplectic structure on $M^{4n}$
by enlarging $M^{4n}$ to superspace $M^{4n|4n}$, with coordinates
$\tilde\lambda^{k} = (\lambda^{a}, \bar c_{i}, c^{j}),\ k, l = 1 \dots 8n$,
endowed with supersymplectic structure defined by
the block diagonal matrix
$(\tilde\Omega^{kl}) = $ diag $(\Omega^{ab}, I^{cd})$,
where $I^{cd}$ is unit $4n \times 4n$ matrix.

\section{$\hbar$-Deformed symplectic structures}\label{QS4}

 Next step is to implement the $\hbar$-deformation.
The $\hbar$-deformed version of the above symplectic structures
is straightforward.
 Namely, we use the Moyal's $\hbar$-deformation\cite{Weyl-Wigner-Moyal}
of the Poisson brackets which plays the role of algebra of quantum
mechanical observables,
\begin{equation}
         \{ f, g\}_{\hbar\omega}      \label{mb}
        = \frac{1}{i\hbar}(f*g-g*f)
        = f\frac{2}{\hbar}
        \sin (\frac{\hbar}{2}\bar\partial_{i}\omega^{ij}\vec\partial_{j})g,
\end{equation}
where the Moyal product is
$
f*g = f\exp (\frac{i\hbar}{2}\bar\partial_{i}\omega^{ij}\vec\partial_{j})g
$.
In the classical limit, the $\hbar$-deformed product and brackets
cover the usual pointwise product,
$
f*g = fg + O(\hbar)
$,
and the Poisson brackets,
$
\{ f, g\}_{\hbar\omega} = \{ f, g\}_{\omega} + O(\hbar^{2})
$,
respectively.

 The Lie-derivatives along the Hamiltonian vector field,
$\ell_{h} = h^{i}\partial_{i}$,
being linear maps, obey the conventional commutation relation,
$
[\ell_{h_{1}},\ell_{h_{2}}] = \ell_{[ h_{1},h_{2}]}
$,
{\sl i.e.} form a Lie algebra.
 The underlying algebra of Hamiltonian vector fields,
$
\bigl[ h_{f}, h_{g} \bigr] = h_{\{ f,g \}_{\hbar\omega}}
$,
is also a Lie algebra due to anticommutativity of
the $\hbar$-deformed brackets (\ref{mb}).
 Quantum mechanical properties of the theory are thus
encoded in these brackets, and the $\hbar$-deformation
preserves the Lie algebra structure of the classical
formalism, with the standard Lie-Poisson algebra replaced by
the Lie-Moyal algebra.

 Our aim is to exploit the symplectic structure $\Omega$ on
$M^{4n}$ introduced above which appears to be crucial
in finding the $\hbar$-deformed Hamilton function ${\cal H}_{\hbar}$.

 Following the definition of the extended Moyal brackets\cite{GR-92d},
we use the symplectic structure on $M^{4n}$ to define
the $\hbar$-deformed Poisson brackets,
\begin{equation}
 \{ F, G\}_{\hbar\Omega} =    \label{emb}
    F\frac{2}{\hbar}
 \sin (\frac{\hbar}{2}\bar\partial_{a}\Omega^{ab}\vec\partial_{b})G,
\end{equation}
with the underlying $\hbar$-deformed product
$
F*G = F\exp (\frac{i\hbar}{2}\bar\partial_{a}\Omega^{ab}\vec\partial_{b})G.
$
 Then, under the $\hbar$-deformation the sequence of maps (\ref{maps})
remains the same, with the Poisson brackets replaced by
the $\hbar$-deformed Poisson brackets (\ref{mb}) and  (\ref{emb}),
respectively.
 Since the $\hbar$-deformed product of functions,
$f(a)*g(a)$ and $F(\lambda)*G(\lambda)$, is non-commutative,
we deal in fact with {\sl non-commutative} cotangent fiber bundles
$M^{2n}$ and $M^{4n}$, which can be studied in terms
of non-commutative geometry\cite{Connes-93}.

 Note that by rescaling $q_{i}$ one has
$\{ F, G\}_{\hbar\Omega} =  \{ F_{\hbar}, G_{\hbar}\}_{1\Omega}$,
where $F_{\hbar}(a,q) \equiv F(a,\hbar q)$,
so that the deformation parameter $\hbar$ can be assigned, equivalently,
to functions on $M^{4n}$ instead of the brackets.
 With the aid of the Grassmannian piece, the brackets (\ref{emb})
become the graded brackets
\begin{equation}
\{ F, G\}_{\hbar\tilde\Omega} =    \label{semb}
       F\frac{2}{\hbar}
  \sin (\frac{\hbar}{2}\bar\partial_{k}\tilde\Omega^{kl}\vec\partial_{l})G,
\end{equation}
where
$
\partial_{k} =\partial /\partial\tilde\lambda^{k}
$.
 Here, the functions $F, G, \dots $ are defined on the superspace
$M^{4n|4n}$, and correspond to antisymmetric tensor fields and
exterior forms on $M^{2n}$, which are (candidates to)
observables of the theory.

 The commutator of the Lie-derivatives on $M^{2n}$,
together with the underlying $\hbar$-deformed algebra
of Hamiltonian vector fields,
can be represented as the $\hbar$-deformed Poisson brackets on $M^{4n}$,
$
     \bigl[ \ell_{h_{1}},\ell_{h_{2}}\bigr]
            \leftrightarrow
\{ \ell_{h_{1}},\ell_{h_{2}}\}_{\hbar\tilde\Omega}
$,
for {\sl horizontal} Hamiltonian vector fields on $M^{4n}$,
$h^{i}(\lambda )$, {\sl i.e.} the fields orthogonal to the
fibers $T^{*}_{a}M^{2n}$.

 Due to coexistence of two symplectic structures,
$(M^{2n}, \omega )$ and $(M^{4n}, \Omega )$, the main point is
to provide consistency between them.
 We require symplectic diffeomorphisms
of the bundle $(M^{4n}, \Omega )$ to preserve symplectic structure on
the base $(M^{2n}, \omega )$.
 That is, under the natural projection,
(i)
$p: \{ , \}_{\hbar\tilde\Omega} \rightarrow \{ , \}_{\hbar\omega}$
and
(ii)
$p: h^{a}(\lambda ) \rightarrow h^{i}(a)$.
 Here, the Hamiltonian vector field on $M^{4n}$ is
$h^{a}(\lambda ) = \Omega^{ab}\partial_{b}H(\lambda )$
so that
$p: H(\lambda ) \rightarrow H(a)$.

 The condition (i) in the form
$
p: \{ {\cal H}_{\hbar}, \rho (a) \}_{1\Omega}
\rightarrow
\{ H(a), \rho (a) \}_{\hbar\omega}
$
has been solved\cite{GR-92d} to find the ghost-free part of
${\cal H}_{\hbar}$, where the projection $p$
provides so called {\sl horizontal condition},
$
q_{i} = \bar c_{i} = c^{i} = 0.
$
 We see that this condition naturally arises from the
consistency requirements (i)-(ii).
 The ghost-dependent part of ${\cal H}_{\hbar}$
is fixed uniquely due to the BRST invariance.
 Namely, the result is (cf.\cite{GR-92d})
\begin{equation}
                                           \label{H1}
{\cal H}_{\hbar}(q,a) = q_{i}h^{i}_{\hbar}
              + i\bar c_{i} \partial_{k}h^{i}_{\hbar}c^{k},
\end{equation}
where
\begin{equation}
H_{\hbar}(q,a) = \frac{f(x)}{x}H(a)                       \label{Hh}
               \equiv \int^{1}_{-1} du \exp [-\hbar xu]H(a)
               \equiv \int^{1}_{-1} du H(a^i - \hbar q_j\omega^{ij}u)
\end{equation}
is $\hbar$-deformed classical Hamiltonian,
$f(x) = \mbox{sh}(x)$,
$ x = \hbar q_{i}\omega^{ij}\partial_{j}$,
and
$
h^{i}_{\hbar} = \omega^{ij}\partial_{j}H_{\hbar}(q,a)
$
is $\hbar$-deformed Hamiltonian vector field.

 The Hamilton function (\ref{H1}) is explicitly $\hbar$-deformed
version of the Hamilton function (\ref{H}),
and plays the same role in the cohomological quantum mechanics
as ${\cal H}$ in cohomological classical
mechanics\cite{Aringazin-95-HJ1}.
 It can be readely verified that in the ghost-free part it
reproduces Wigner operator\cite{Aringazin-94-JP}.

 Generally, the states are defined by the $p$-ghost Wigner density
$\rho=\rho(a,c,t)$, and the flow equation is
$\partial_{t}\rho = \{ \rho, H\}_{\hbar\omega} = {\cal H}_{\hbar}\rho$.
 In the classical limit, $\hbar \rightarrow 0$,
the Hamilton function (\ref{H1}) reduces to the Hamilton function
(\ref{H}), and the flow equation reduces to the conventional
Liouville equation, in the ghost-free part.

 Due to the (anti-)BRST symmetry of the underlying classical theory
(\ref{H}), we should study the associated symmetry of the
$\hbar$-deformed Hamilton function (\ref{H1}).
 The only difference from the classical case may arise from the BRST
transformation of $h^{i}_{\hbar}$.
 Namely,
$\delta h^{i}_{\hbar} = \partial h^{i}_{\hbar}/\partial a^k\delta a^k
                      + \partial h^{i}_{\hbar}/\partial q_k\delta q_k$,
and since $sa^i = c^i$, $sq_i = 0$, we have
$s h^{i}_{\hbar} = \partial_k h^{i}_{\hbar} c^k$.
 This means that the BRST symmetry survives the $\hbar$-deformation,
and ${cal H}_\hbar$ is BRST invariant, $s{\cal H}_\hbar = 0$.

\end{document}